\documentclass[10pt,conference]{IEEEtran}

\newcounter{finding}

\usepackage{graphicx}
\usepackage{multirow}
\usepackage{amsthm}

\usepackage{color}
\usepackage{tcolorbox}
\usepackage{algorithm}
\usepackage{algorithmic}
\usepackage{footmisc}
\usepackage{caption}
\usepackage{amsmath}
\usepackage{xspace}
\usepackage{hyperref}
\usepackage{cleveref}
\usepackage{booktabs}
\usepackage{wrapfig}
\usepackage{pifont}
\usepackage{subfigure}
\usepackage{mathbbol}
\usepackage{mdframed}
\usepackage{tikz}
\usepackage{colortbl}
\usepackage{listings}
\usepackage{xcolor}

\lstdefinestyle{pythonstyle}{
    language=Python,
    basicstyle=\ttfamily\scriptsize,
    keywordstyle=\color{blue}\bfseries,
    commentstyle=\color{green!60!black},
    stringstyle=\color{red},
    numberstyle=\tiny\color{gray},
    numbers=left,
    numbersep=5pt,
    frame=single,
    breaklines=true,
    showstringspaces=false,
    tabsize=2
}

\usepackage{url}

\usepackage{fancyhdr}
\def\Snospace~{\S{}}

\renewcommand{\figurename}{Fig.}

\usepackage{makecell}
\usepackage{gensymb}
\newcounter{num}
\usepackage{diagbox}
\newcommand{\find}[1]{ \begin{tcolorbox}
\textbf{Finding \refstepcounter{num}\thenum}: #1
\end{tcolorbox}}

\begin{document}

\title{The Foundation Cracks: A Comprehensive Study on Bugs and Testing Practices in LLM Libraries}

\author{
    \IEEEauthorblockN{Weipeng Jiang$^{1}$, Xiaoyu Zhang$^{1,4}$, Xiaofei Xie$^{2}$, Jiongchi Yu$^{2}$, Yuhan Zhi$^{1}$, Shiqing Ma$^{3}$, Chao Shen$^{1}$}
    \IEEEauthorblockA{$^1$ Xi'an Jiaotong University, China $^2$ Singapore Management University, Singapore}
    \IEEEauthorblockA{$^3$ University of Massachusetts Amherst, USA $^4$ Nanyang Technological University, Singapore
     }
    
    \IEEEauthorblockA{\textit{\{lenijwp, zxy0927, zyh1123\}@stu.xjtu.edu.cn, jcyu.2022@phdcs.smu.edu.sg}} 
    \IEEEauthorblockA{\textit{xfxie@smu.edu.sg, shiqingma@umass.edu,  chaoshen@xjtu.edu.cn}}
    \vspace{-2.1em} 
}

\maketitle

\begin{abstract}
Large Language Model (LLM) libraries have emerged as the foundational infrastructure powering today's AI revolution, serving as the backbone for LLM deployment, inference optimization, fine-tuning, and production serving across diverse applications.
Despite their critical role in the LLM ecosystem, these libraries face frequent quality issues and bugs that threaten the reliability of AI systems built upon them.
To address this knowledge gap, we present the first comprehensive empirical investigation into bug characteristics and testing practices in modern LLM libraries.
We examine 313 bug-fixing commits extracted across two widely-adopted LLM libraries: HuggingFace Transformers and vLLM.
Through rigorous manual analysis, we establish comprehensive taxonomies categorizing bug symptoms into 5 types and root causes into 14 distinct categories.
Our primary discovery shows that API misuse has emerged as the predominant root cause (32.17\%-48.19\%), representing a notable transition from algorithm-focused defects in conventional deep learning frameworks toward interface-oriented problems.
Additionally, we examine 7,748 test functions to identify 7 distinct test oracle categories employed in current testing approaches, with predefined expected outputs (such as specific tensors and text strings) being the most common strategy.
Our assessment of existing testing effectiveness demonstrates that the majority of bugs escape detection due to inadequate test cases (41.73\%), lack of test drivers (32.37\%), and weak test oracles (25.90\%).
Drawing from these findings, we offer some recommendations for enhancing LLM library quality assurance.
\end{abstract}



\section{INTRODUCTION}\label{sec:intro}

The emergence of Large Language Models (LLMs) as a dominant computing paradigm has fundamentally reshaped AI development practices, revealing that traditional deep learning frameworks like PyTorch~\cite{paszke2019pytorch} and TensorFlow~\cite{tensorflow2015-whitepaper}, while effective for general machine learning tasks, cannot adequately support the unique requirements of large-scale LLM applications. In contrast, specialized LLM libraries including HuggingFace Transformers~\cite{wolf-etal-2020-transformers}, vLLM~\cite{kwon2023efficient}, and DeepSpeed~\cite{10.1145/3394486.3406703} have gained widespread adoption by offering distinctive advantages: comprehensive model ecosystems with unified APIs for seamless model access and management, advanced memory optimization techniques such as PagedAttention~\cite{kwon2023efficient} that dramatically reduce resource consumption, and integrated end-to-end workflows spanning from model fine-tuning to production serving. These innovations have positioned such libraries as the foundational infrastructure for modern LLM development.

In the past few years, the rapid evolution of LLM technologies has demanded frequent updates and fast iteration for LLM libraries to support emerging model architectures, advanced inference techniques, and novel training methodologies~\cite{zhu2024architecturalfoundationslargelanguage,duan2024efficienttraininglargelanguage}. This accelerated development pace raises significant concerns about the quality assurance of these critical software systems.
Unfortunately, real-world evidence suggests that numerous bugs and defects have indeed emerged across the LLM library ecosystem. 
For instance, security vulnerabilities in HuggingFace Transformers model loading mechanisms have enabled model poisoning attacks in production environments~\cite{realcase_2024_bytedance}, while incorrect implementations of gradient accumulation across multiple LLM libraries have silently compromised the training quality of countless models over recent years~\cite{realcase_2024_reddit}. These incidents highlight the urgent need for a systematic investigation into LLM library bugs and seek for better test practices to ensure their software quality.

Previous work has extensively studied bug characteristics and testing practices across various traditional DL frameworks~\cite{10.1145/3338906.3338955,chen2023toward,cao2021characterizing,quan2022towards}. 
The SE community has also actively developed various testing methods for DL libraries to discover and help repair bugs~\cite{deng2023largelanguagemodelszeroshot, 10.1109/ICSE48619.2023.00017,10.1145/3575693.3575707}. 
However, to date, there remains a significant knowledge gap regarding the in-depth understanding of bugs in LLM libraries and effective testing strategies specifically for these new complex software systems.
While traditional DL frameworks serve as foundational computational platforms, LLM libraries are built atop these frameworks to provide more domain-specific functionalities tailored for LLM workflows. This architectural distinction raises critical questions about whether existing empirical findings and testing methodologies from traditional DL frameworks are applicable to LLM libraries.
Therefore, there is an urgent need for dedicated empirical investigation to understand the distinctive quality assurance challenges in the LLM library ecosystem and develop appropriate testing strategies.

To fill this knowledge gap, we conduct the first comprehensive empirical study about bug characteristics and testing practices in modern LLM libraries, aiming to answers three key research questions:
\begin{itemize}
\item \textbf{RQ1 (Bug Symptoms)}: What are the symptoms of bugs in LLM libraries and their distributions?

\item \textbf{RQ2 (Root Cause)}: What are the underlying root causes of bugs in LLM libraries, and how are they distributed?

\item \textbf{RQ3 (Testing Practices)}: What testing strategies are currently employed in LLM libraries, and how effective are they in detecting real-world defects?

\end{itemize}

\begin{figure*}[!t]
    \centering     
    \includegraphics[width=0.95\linewidth]{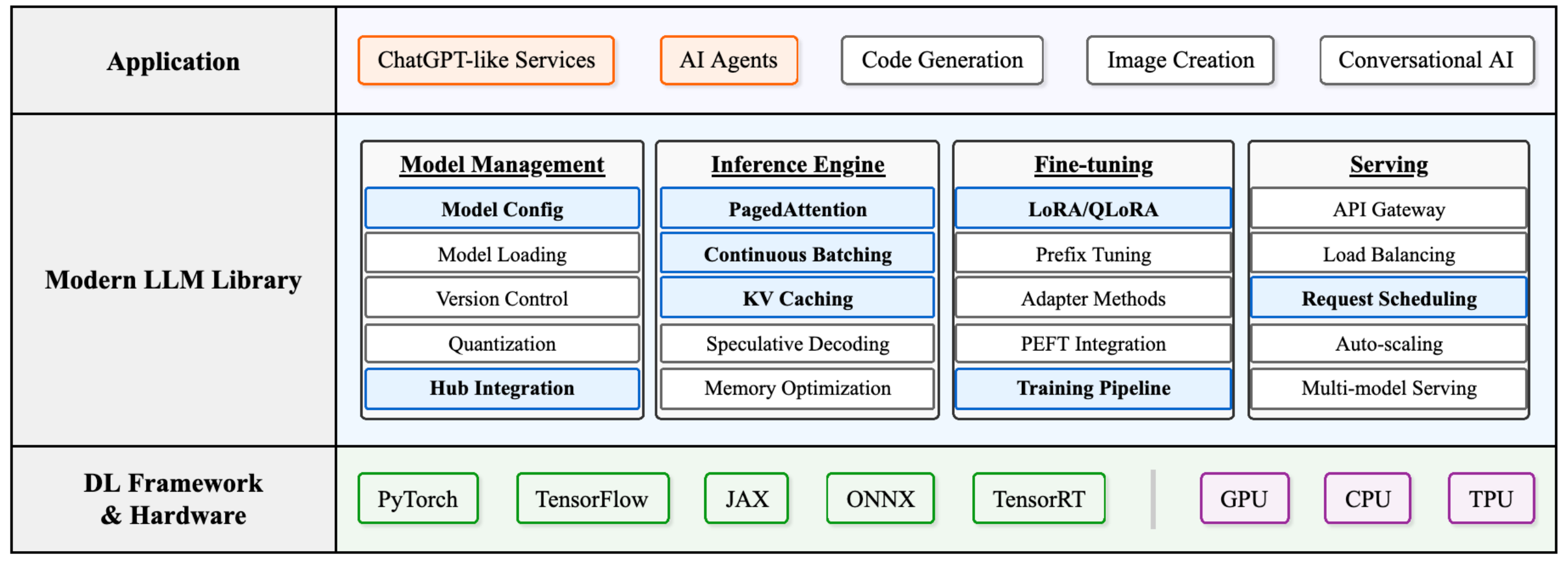}
    \caption{Typical Architecture of Modern LLM Libraries}
    \label{fig:lib_structure}
    \vspace{-5pt}
\end{figure*}

To answer these research questions, this study focuses on two representative and widely adopted LLM libraries: HuggingFace Transformers~\cite{wolf-etal-2020-transformers} (the foundational ecosystem backbone with over 144,000 GitHub stars) and vLLM~\cite{kwon2023efficient} (a leading high-performance inference engine with 48,600 GitHub stars). 
We systematically collect and analyze 313 bug-fixing commits extracted from 6,204 total commits over a 12-month period through rigorous keyword-based filtering followed by manual validation to ensure scientific rigor. The analysis further examines 7,748 test functions to comprehensively characterize current testing practices. To evaluate testing effectiveness, the study executes these test suites against the collected bug-fix commits to determine their capability in detecting real-world failures.

Through systematic analysis, our study develops comprehensive taxonomies comprising 5 distinct bug symptom categories and 14 root cause categories to characterize the bug landscape in LLM libraries. 
The investigation further categorizes existing testing practices into three granularity levels and identifies 7 different test oracle strategies employed by current test practices. Notably, the analysis reveals that the majority of collected bugs cannot be detected by contemporary test suites, with fine-grained analysis identifying three primary categories of testing inadequacies.
Our study finds that:
\ding{182}
Crashes and incorrect functionalities remain the most prevalent symptoms across LLM libraries and deep learning (DL) frameworks.
\ding{183}
API misuse has emerged as the dominant root cause, accounting for 32.17\%-48.19\% of all bugs, marking a significant departure from traditional DL frameworks~\cite{chen2020comprehensive} where algorithm implementation errors typically dominate. The rise of API misuse reflects the higher level of abstraction and the intricate interdependencies among components in modern LLM libraries.
\ding{184}
Testing in existing LLM libraries primarily relies on manually defined test oracles (e.g., handwritten expected outputs), which account for approximately 40.48\% of all tests. However, this reliance raises concerns about the sufficiency and extensibility of current testing practices, particularly given the dynamic behavior of LLMs and the increasing diversity of usage scenarios.
\ding{185}
Most bugs remain undetected by the existing test suites, primarily due to three main reasons: lack of test drivers (32.37\%), test cases (41.73\%), test oracles (25.90\%). This highlights significant gaps in current testing practices and the need for more effective testing strategies.

In summary, this paper makes the following key contributions:
\begin{itemize}
    \item To the best of our knowledge, this is the first empirical study to systematically explore bug characteristics and test practices in modern LLM libraries. 
    \item We provide a series of findings and actionable insights that benefit multiple stakeholders, including LLM library developers, users, and test researchers.
    \item We release our dataset of 313 bug-fixing commits and 7,748 test cases~\cite{ourrepo}, along with the detailed analysis results, to facilitate further research in this area.
\end{itemize}


\section{Background}
\label{sec:background}

With the rapid advancement of LLMs, traditional deep learning (DL) frameworks have proven inadequate for handling the increasing scale and complexity of modern language models. Specifically, they lack efficient support for memory-intensive inference, advanced request scheduling, and high-concurrency deployment scenarios. Consequently, modern LLM libraries have emerged to address these evolving requirements. As illustrated in~\autoref{fig:lib_structure}, these libraries provide a comprehensive framework built upon four core components:

\textbf{\textit{Model Management}} serves as the foundation for LLM ecosystems, providing unified model configuration and efficient loading mechanisms for multi-gigabyte weights across distributed storage. Key capabilities include version control for model iteration tracking, quantization techniques for memory footprint reduction, and seamless hub integration with model repositories like HuggingFace, enabling practitioners to easily access and deploy diverse pre-trained models.

\textbf{\textit{Inference Engine}} is tailored to the computational demands of LLMs and includes several core optimizations.
PagedAttention models attention computation as virtual memory paging to mitigate memory fragmentation, while continuous batching supports dynamic request processing to maximize GPU utilization. KV caching reduces redundant computation during autoregressive decoding, and speculative decoding accelerates inference by pre-generating tokens, collectively contributing to significant improvements in throughput and latency.
    
\textbf{\textit{Fine-tuning}} enables efficient model adaptation through parameter-efficient techniques like LoRA~\cite{hu2022lora}/QLoRA~\cite{dettmers2023qlora}  with minimal computational overhead. Integrated training pipelines with PEFT support orchestrate the complete customization workflow, allowing practitioners to adapt models for specific tasks without full retraining.
    
\textbf{\textit{Serving}} provides production-ready deployment infrastructure with sophisticated request scheduling algorithms, intelligent load balancing across multiple instances, auto-scaling capabilities for dynamic resource management, and multi-model serving support for diverse application scenarios.

These layered components work synergistically, building upon existing DL frameworks (e.g., PyTorch, TensorFlow, JAX) and leveraging heterogeneous hardware backends (e/g., GPU, CPU, TPU) through optimized runtime engines, creating a specialized ecosystem tailored for modern LLMs.

\section{STUDY METHODOLOGY}\label{sec:method}

\subsection{Data Collection}\label{sec:method_collection}

\subsubsection{Target Libraries Selection}\label{sec:method_collection_target}

In this study, we focus on two modern LLM libraries: \textbf{HuggingFace Transformers} (referred to as Transformers) and \textbf{vLLM}.
Transformers serves as the foundational backbone of today’s LLM ecosystem, with over 144,000 GitHub stars and more than 500,000 hosted pre-trained models, supporting a wide range of applications from training to deployment~\cite{wolf-etal-2020-transformers}.
vLLM, with 48,600 GitHub stars, represents a high-performance inference engine originally developed at UC Berkeley~\cite{kwon2023efficient}. Known for innovations like PagedAttention and excellent throughput, it has quickly evolved into a community-driven project bridging research and production.
These two libraries offer complementary perspectives.
Transformers reflects general-purpose LLM development and model management, while vLLM focuses on inference efficiency and runtime optimization.
Our selection of these two libraries is primarily driven by two key factors: \ding{182} their widespread adoption in both academia and industry, and \ding{183} the strong support they receive from professional communities and dedicated maintenance teams.
Crucially, both libraries provide comprehensive test suites that help their quality assurance.
This aspect is vital, as it directly enables us to conduct a meaningful analysis of industry-standard testing practices and helps to gain a deep understanding of quality assurance methods within the LLM software ecosystem.


\subsubsection{Bug Collection}\label{sec:method_collection_bug}

Following previous works~\cite{yu2024bugs,quan2022towards,shi2022large}, we collect and analyze bug-fixing commits from the Transformers and vLLM Github repositories.
The reasons why using such bug-fixing commits are twofold: \ding{182} All bug-related issues are typically linked to corresponding fix commits, while some bugs are fixed without associated issue reports (e.g., independently discovered and fixed by developers).
\ding{183} Bug-fixing commits allow us to localize buggy code precisely, enabling a finer-grained analysis of the root cause and whether the corresponding test cases can detect these bugs.

We collect and analyze the commits over a 12-month period from January 2024 to December 2024.
In total, we collect 2,850 and 3,354 commits from Transformers and vLLM, respectively.
We then apply a two-stage filtering process to identify genuine bug-fix commits:
\ding{182} \textbf{Automatic Filtering (AF)}, We first perform the automatic filtering (AF). 
Following established practices~\cite{shen2021comprehensive, yu2024bugs}, we select commits whose messages contain bug-related keywords (e.g., \textit{fix}, \textit{defect}, \textit{error}, \textit{bug}, \textit{issue}, \textit{mistake}, \textit{correct}, \textit{fault}, \textit{flaw}).
We remove duplicates such as merge commits (e.g., messages containing \textit{merge} or \textit{PR}) and ensure that the test status of the commit is recorded as \texttt{success}.
This step yields 992 and 357 candidate commits for Transformers and vLLM, respectively.
\ding{183} \textbf{Manual Filtering (MF)} We further exclude commits that only involve test fixes or non-functional changes like documentation corrections.
After manual inspection, we identify 230 and 83 valid bug-fix commits for Transformers and vLLM, respectively—313 bug-fix commits in total for our analysis, as summarized in~\autoref{tab:llm_issues}.

\begin{table}[!tbp]
    \centering
    \caption{Collected Bug-fixing Commits}
    \label{tab:llm_issues}
    \begin{tabular}{l|c|c|c}
    \hline
    \textbf{Repository Type}  & \textbf{Commits Crawled} & \textbf{After AF} & \textbf{After MF} \\
    \hline
    \textbf{Transformers}                            & 2,850                        & 992             & 230             \\
    \textbf{vLLM}                                       & 3,354                        & 357             & 83             \\
    \hline
    \textbf{Total}                    & \textbf{6,204}            & \textbf{1,349} & \textbf{313} \\
    \hline
    \end{tabular}
\end{table}

\subsubsection{Test Collection}\label{sec:method_collection_test}

To understand the current testing practices adopted by modern LLM libraries and to investigate the limitations of these practices in detecting real-world bugs, we analyze the existing test suites of the selected libraries.
Specifically, we collected the test suites from the latest stable releases of HuggingFace Transformers and vLLM during our study period. These suites generally fall into two major categories: unit tests and integration tests.

In the initial collection phase, we extracted all unique test functions, resulting in 9,031 test functions for Transformers and 862 for vLLM.
To avoid overcounting, we excluded inherited test functions where a subclass simply invokes a parent class’s test method. These duplicates do not introduce new testing logic or coverage and would otherwise distort our analysis.
Next, we performed a second-round filtering to exclude non-implemented or placeholder tests. We observed a substantial number of test functions in both libraries that are either marked as not yet implemented or intentionally left as placeholders with only \(\textit{Pass}\). This is often due to the corresponding features still being under development or uncertainty about how to design meaningful test cases for them at the time.
After filtering out such cases, we retained 6,897 and 851 (7,748 in total) valid test functions for Transformers and vLLM, respectively.
This refined dataset forms the basis of our analysis in understanding how developers of modern LLM libraries currently test their systems and the extent to which these tests can detect real-world bugs.

\subsection{Classification and Labeling Process}\label{sec:method_labeling}

This study employs a systematic multi-phase classification methodology to ensure scientific rigor and reproducibility in bug and test characterization.

\subsubsection{Taxonomy Development}\label{sec:method_labeling_taxonomy}
The classification framework was developed through an iterative grounded theory approach. 
Initial taxonomies were grounded in established frameworks from prior DL library studies~\cite{chen2020comprehensive}, then systematically extended through open coding of LLM-specific patterns. 
For bug-fix commits, the analysis characterized both manifestation symptoms and underlying root causes through careful examination of code changes, issue reports, and commit messages. For test cases, a dual-dimensional taxonomy was constructed encompassing granularity levels and oracle mechanisms.

\subsubsection{Multi-Rater Annotation Protocol} 
To ensure classification reliability, the study implemented a rigorous multi-phase annotation protocol following established qualitative research standards~\cite{landis1977measurement}. 
Two independent annotators with expertise in software testing performed initial coding on a stratified sample (20\% of the dataset). Inter-rater reliability was measured using Cohen's Kappa coefficient, with disagreements systematically resolved through structured discussion and consultation with a senior arbitrator. The taxonomy was iteratively refined until achieving substantial agreement (Cohen’s Kappa coefficient exceeding 0.80), at which point the finalized schema was applied to the complete dataset.

\subsubsection{Testing Effectiveness Analysis} 
To empirically evaluate current testing practices, the study executed collected test suites against the bug-fix commits to obtain precise coverage metrics. This execution-based analysis enabled systematic identification of testing gaps by cross-referencing coverage with known bugs, providing quantitative evidence for the three categories of testing inadequacies identified in our findings.

\section{RQ1: Bug Symptoms}\label{sec:rq1}

\subsection{Symptom Classification Results}\label{sec:rq1_symptom}


\begin{figure}[!tbp]
    \centering     
    \includegraphics[width=0.9\linewidth]{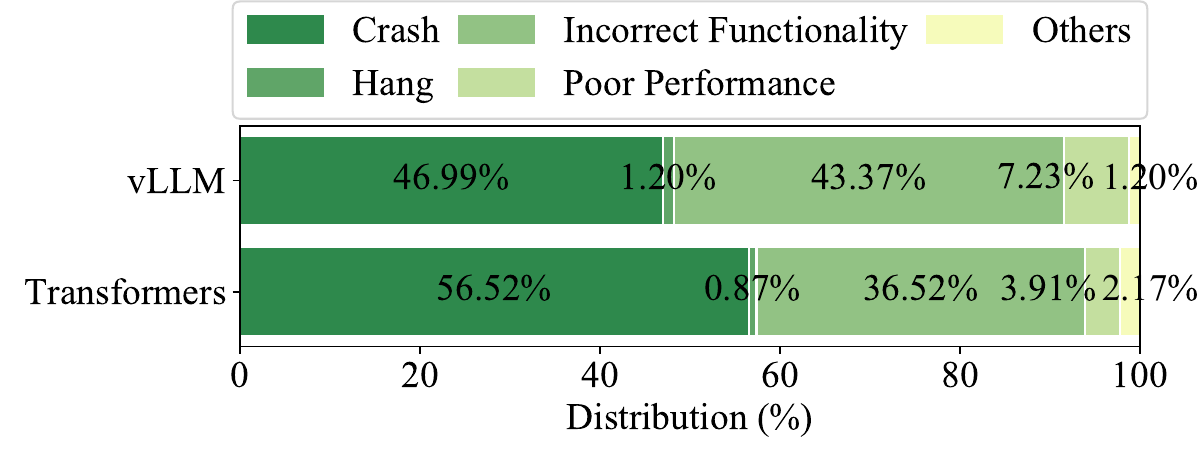}
    \caption{Bug Distribution by Symptoms}
    \label{fig:symptom}
\end{figure}

\subsubsection{Crash}
This symptom refers to situations where LLM libraries terminate unexpectedly during execution, typically accompanied by error messages such as `TypeError'.
For example, an issue in the Transformers library reports that using specific input dimensions (e.g., 384x384) in \texttt{SiglipVisionModel.forward} can lead to a crash, raising a TypeError~\cite{transformers_issue33993}.
An issue report in vLLM states that when using the Mistral model, the program might crash because the \texttt{SentencePieceTokenizer} lacks the attribute `id\_to\_byte\_piece'~\cite{vllm_issue9907}.

\subsubsection{Incorrect Functionality}
This symptom refers to situations where LLM libraries execute successfully without crashing but produce incorrect or unexpected results.
Such silent failures are particularly dangerous, as they often go undetected in production environments and may result in downstream misbehaviors.
For instance, an issue report in the vLLM project~\cite{vllm_pull10903} highlights that, due to the \texttt{max\_model\_len} variable not being divisible by the \texttt{block\_size} variable in the underlying implementation, the \texttt{llm.generate()} function may produce incorrect outputs in specific cases.

\subsubsection{Poor Performance}
This symptom refers to scenarios where the time or computational resources (e.g., memory) consumed during the use of an LLM library significantly exceed expected levels, leading to degraded system performance and limited scalability.
For instance, in the Transformers library, when downstream users invoke the \texttt{get\_state\_dict\_from\_offload} function, a copy of the original GPU tensor is retained longer than anticipated, resulting in substantial memory overhead. In one observed instance, memory usage increased from an expected value of less than 68.9\,GB to over 80\,GB~\cite{transformers_pull34890}.

\subsubsection{Hang}
This symptom refers to scenarios in which programs built on LLM libraries become unresponsive or enter infinite loops, failing to terminate within a reasonable time.
Such behavior raises critical reliability concerns, especially in production systems.
A representative example arises in the Transformers library, which internally relies on \texttt{torch.tensor()} to convert NumPy arrays into PyTorch tensors, a process that involves memory copying~\cite{transformers_issue33185}.
This implementation may cause the program to hang when multiple threads concurrently invoke the library, potentially resulting in deadlocks in multi-threaded execution environments.

\subsubsection{Others}
This symptom refers to other rare symptoms that are not covered by the above symptoms (e.g., build failure, confusing variable naming) or cannot be clearly determined even after reviewing relevant information.
For example, an issue report in the Transformers~\cite{transformers_pull31436} points out that the naming around rope scaling type is confusing and needs to be corrected according to the related literature.

\subsection{Symptom Distributions}\label{sec:rq1_distribution}

\autoref{fig:symptom} illustrates the distribution of bug symptoms in the LLM library.
The most prevalent symptom observed is `Crash', constituting 56.52\% of the bugs associated with 230 commits in Transformers and 46.99\% of the bugs associated with 83 commits in vLLM.
This is followed by `Incorrect Functionality', representing 36.52\% in Transformers and 43.37\% in vLLM.
Other less frequent symptoms include  `Hang' at around 0.87\%–1.20\% and `Poor Performance' at approximately 3.91\%–7.23\%.
The `Others' category accounts for a small fraction, less than 2.17\% of the bugs on two libraries.

Compared with the distribution of bug symptoms observed in DL frameworks reported in prior work~\cite{chen2023toward}, `Crash' continues to dominate in LLM libraries (46.99\%-56.52\%), which is similar to that in DL frameworks (42.40\%-57.60\%).
A significant divergence lies in `Build Failure', while this symptom ranks third in DL frameworks, occupying 11.20\%-28.40\% of bugs, it is almost negligible in LLM libraries, contributing no more than 2.17\% and being subsumed under the `Others' category.
This discrepancy is plausibly attributable to the fact that LLM packages are typically installed via package managers (e.g., \texttt{pip install}) rather than manual compilation, thereby reducing the opportunity for build errors.
Finally, the share of \textit{Poor Performance} in LLM libraries (3.91\%-7.23\%) exceeds that in DL frameworks (1.20\%-3.60\%).
This is likely because LLM libraries embed intricate memory management and sampling algorithms tightly coupled with model inference, rendering them more susceptible to performance bugs.

\find{
Crashes remain the most prevalent symptom across LLM libraries and DL frameworks.
However, compared with DL frameworks, LLM libraries exhibit two distinctive characteristics:
\ding{182} Build failures have largely vanished, and \ding{183} performance-related issues occur proportionally more often.
}


\section{RQ2: Root Causes}\label{sec:rq2}

\subsection{Root Cause Classification Results}\label{sec:rq2_rootcause}

Based on our empirical analysis, we categorize the root causes of bugs observed in modern LLM libraries into 14 distinct categories.
Each category represents a fundamental type of underlying issue that frequently arises during the development, deployment, and maintenance of LLM libraries.
This categorization provides valuable insights into the unique reliability challenges inherent in this rapidly evolving LLM software ecosystem.

\textbf{API Misuse (A).} This category encompasses defects resulting from the incorrect or inappropriate use of LLM library APIs.
These issues often arise due to complex parameter interdependencies and context-sensitive behaviors inherent to LLM APIs.
We identify four primary forms of API misuse.
\ding{182} \textit{Incorrect parameters}, where developers invoke the correct API but pass inappropriate values, such as non-existent parameters or incorrect assignments.
This is often encountered in nuanced configuration or tokenization settings.
\ding{183}\textit{Missing or redundant conditions}, involving omitted or unnecessarily duplicated validation checks prior to API invocation, leading to unpredictable behaviors.
\ding{184} \textit{Wrong API calls}, where syntactically correct but semantically inappropriate APIs are used. 
\ding{185} \textit{Missing or redundant API calls}, where essential API calls are omitted or unnecessary ones are included in the LLM processing pipeline.

\textbf{API Incompatibility (B).} Compatibility issues are a major source of bugs, driven by the rapid evolution of LLM libraries. 
It contains two forms.
\ding{182} \textit{Internal Incompatibility} refers to issues arising from breaking changes across different versions of the same library, often causing regressions.
\ding{183} \textit{External Incompatibility} involves version mismatches and conflicts between LLM libraries and their dependencies, such as tokenizers, model converters, and inference engines.

\textbf{Environment Incompatibility (C).} 
LLM libraries are expected to operate across diverse computational settings, including local development environments and distributed cloud systems.
This root cause captures system-specific conflicts and operational failures related to diverse environmental factors, including hardware acceleration (e.g., GPU models), CUDA compatibility, distributed computing configurations (e.g., multi-node setups), and nuances of containerized deployment environments.

\textbf{Incorrect Algorithm Implementation (D).} This category refers to cases where developers intended to implement a specific algorithm or functionality, but the implementation was incorrect. Such issues often stem from misunderstandings or mistakes in translating algorithmic designs into code. 
For example, a mistake in the placement of a special token like \texttt{[BOS]} can disrupt the input formatting process, ultimately leading to incorrect model inference.


\textbf{Incorrect Exception Handling (E).} Inadequate error handling impedes debugging and obscures critical runtime issues during training or inference.
This includes missing exceptions for invalid states, misleading error messages, and insufficient diagnostic information.

\textbf{Misconfiguration (F).} These bugs result from incorrect settings such as invalid model paths, erroneous initialization files, or improperly specified configuration options.
For example, a misconfiguration may occur when the checkpoint path for a specific model, such as LLaVA, is missing specified~\cite{transformers_pull32458}.

\textbf{Numerical Issues (G).} This category covers problems related to numerical instability that compromise the accuracy or stability of LLM computations.
Specific examples include division by zero errors during attention computations, overflow or underflow conditions during sensitive gradient calculations, and the use of incorrect mathematical operators within loss functions or optimization steps.

\textbf{Type Issues (H).} Type-related problems arise from the use of specialized tensor types and strict precision constraints in LLM operations.
This category encompasses tensor-specific type issues (e.g., incompatible mixed-precision training configurations or attention mask type mismatches) and traditional data type problems affecting model parameters and configuration settings.

\textbf{Tensor Shape Misalignment (I).} 
Shape and dimension mismatches in tensor operations are a frequent and critical issue in LLM systems.
Attention mechanisms and sequence processing pipelines demand precise shape matching across various stages.
LLM libraries must robustly handle variable-length sequences and dynamic attention patterns, making accurate tensor shape alignment during batch processing and model inference especially prone to errors.

\textbf{Concurrency Issues (J).} Bugs in this category relate to erroneous behavior in multi-threaded and distributed processing environments, which are common in LLM training and inference.
These issues include race conditions, improper synchronization of shared computational resources, and inconsistent state management across parallel processes, potentially leading to data corruption or erroneous outputs.

\textbf{Dependent Module Issues (K).} The complex and often evolving dependency landscape of LLM libraries introduces a class of bugs related to external packages.
This includes missing imports, incompatible version combinations, and dependency conflicts that result in model loading failures or runtime errors.

\textbf{API Missingness (L).} This root cause refers to the absence of expected API methods or functions that should be available based on common interface design or usage patterns.
Unlike `API Misuse' (A), which involves incorrect use of existing APIs, this reflects a fundamental gap in functionality.
For instance, a model may lack essential methods such as \texttt{get\_output\_embeddings}, limiting its interoperability within the LLM ecosystem.
Such omissions typically arise during rapid feature development, when supporting new model architectures, or when adherence to a common interface is inadvertently overlooked.

\textbf{Incorrect Assignment or Initialization (M).} This category covers fundamental programming errors related to variable assignment and object initialization.
Examples include improperly initialized model weights at the start of training, incorrect default parameter values affecting model behavior, and the assignment of invalid values to model configuration attributes.



\textbf{Others (N).}
This category includes issues that do not fall under the above classifications.
These may represent rare, highly specific, or potentially emerging bug patterns that warrant further investigation as LLM library development practices continue to evolve.


\subsection{Root Cause Distributions}\label{sec:rq2_distribution}

\begin{figure}[!tbp]
    \centering     
    \includegraphics[width=\linewidth]{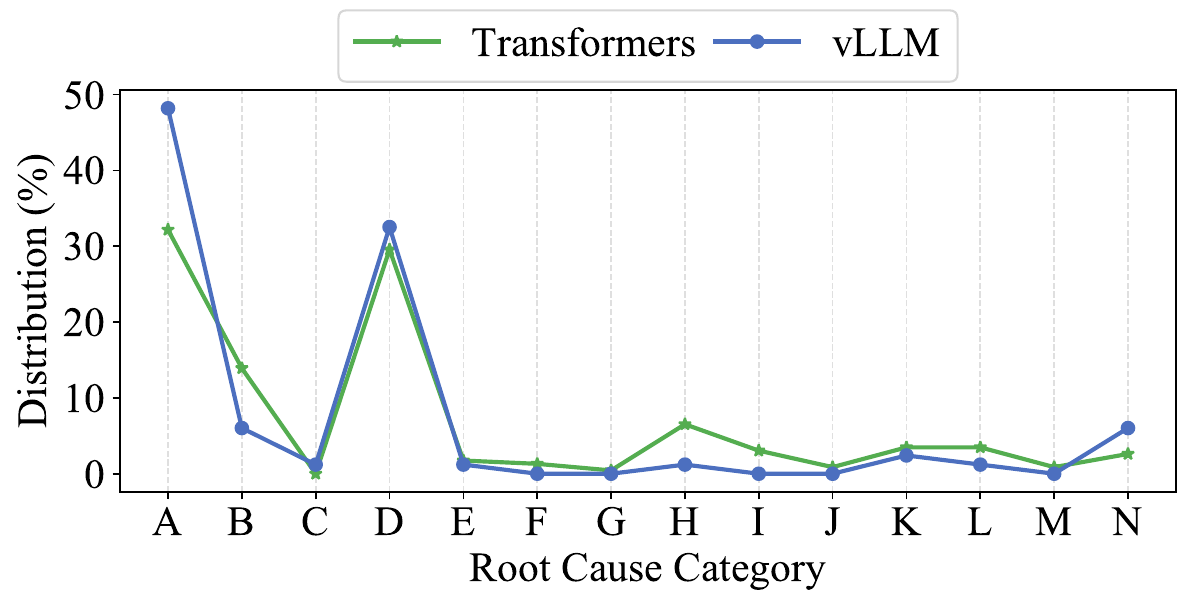}
    \caption{Bug Distribution by Root Causes \footnotesize{(X-axis Shows the Abbreviations of Each Category)}}
    \label{fig:rq2}
    \vspace{-5pt}
\end{figure}

\autoref{fig:rq2} shows the distribution of bug root causes for the Transformers (green) and vLLM (blue) libraries. 
We can observe that the two libraries have similar root cause distributions.
Specifically, in both LLM libraries, the dominant category is `API Misuse (A)', reaching 32.17\% in Transformers and 48.19\% in vLLM.
The runner-up is `Incorrect Algorithm Implementation (D)', accounting for 29.57\% and 32.53\%, respectively, so that these two categories have explained well over half of all collected bugs.
Mid-tier contributors differ: Transformers shows a visible share of `API Incompatibility (B)' at 13.91\% whereas vLLM registers only 6.02\%.
Conversely, vLLM has a heavier ratio in `Others (N)', hinting at several nascent, project-specific bug patterns.

Compared with DL frameworks, where `Incorrect Algorithm Implementation (D)' is the most common root cause, `Type Issues (H)' ranks second, and `API Misuse (A)' ranks fifth~\cite{chen2023toward}, LLM libraries present a markedly different characteristic.
Two factors could help explain the reversal.
First, LLM libraries expose rapidly evolving, fine-grained APIs for tokenization, sampling, and memory-efficient execution.
Developers frequently experiment with the new features, resulting in a surge in bugs with incorrectly specified parameters or calls (A).
In addition, both LLM libraries inherit mature tensor and mixed precision abstractions from DL frameworks, which greatly reduces low-level type mismatches (H)
Therefore, while algorithmic errors (D) are still a major root cause of bugs, API-centric bugs have become a new and important challenge facing LLM libraries.

\find{
The distributions of root causes of the two LLM libraries are similar.
The most common root cause is `API Misuse'.
Compared with DL frameworks, LLM libraries shift the bug landscape from algorithm- and type-centric faults toward interface-centric ones, making rigorous API design and validation an important requirement for LLM library development and maintenance.
}

\begin{table}[!tbp]
    \centering
    \scriptsize
    \caption{Bug Distribution By Symptoms for Each Root Cause \footnotesize{(I.F Refers to Incorrect Functionality, P.P Refers to Poor Performance, and the Columns Refer to the Abbreviations of Root Causes)}}
    \label{tab:rq2}
    \begin{tabular}{c|c|c|c|c|c|c}
    \hline
    \diagbox[height=1cm, width=2.5 cm]{\textbf{Root Cause}}{\textbf{Symptom}} & \textbf{Crash} & \textbf{Hang} & \textbf{I.F} & \textbf{P.P} & \textbf{Others} & \textbf{Total} \\
    \hline
    \textbf{A} & 65 & 1 & 40 & 8 & 0 & 114 \\
    \textbf{B} & 27 & 0 & 8 & 0 & 2 & 37 \\
    \textbf{C} & 0 & 0 & 1 & 0 & 0 & 1 \\
    \textbf{D} & 42 & 1 & 46 & 5 & 1 & 95 \\
    \textbf{E} & 1 & 0 & 3 & 1 & 0 & 5 \\
    \textbf{F} & 3 & 0 & 0 & 0 & 0 & 3 \\
    \textbf{G} & 1 & 0 & 0 & 0 & 0 & 1 \\
    \textbf{H} & 11 & 0 & 4 & 1 & 0 & 16 \\
    \textbf{I} & 6 & 0 & 1 & 0 & 0 & 7 \\
    \textbf{J} & 0 & 1 & 1 & 0 & 0 & 2 \\
    \textbf{K} & 7 & 0 & 1 & 0 & 2 & 10 \\
    \textbf{L} & 3 & 0 & 6 & 0 & 0 & 9 \\
    \textbf{M} & 0 & 0 & 1 & 0 & 1 & 2 \\
    \textbf{N} & 3 & 0 & 8 & 0 & 0 & 11 \\
    \hline
    \textbf{Total} & 169 & 3 & 120 & 15 & 6 & 313 \\
    \hline
    \end{tabular}
\end{table}

\autoref{tab:rq2} shows the distribution of root causes and symptoms for all collected bugs on two LLM libraries.
We can observe that two API-centric root causes, namely `API Misuse' and `API Incompatibility', are highly likely to cause crashes.
They alone cause 92 of the 169 recorded crashes, reflecting the serious harm of incorrectly specified parameters or version mismatches to the development, which often leads to immediate runtime failures.
In addition, `Incorrect Algorithm Implementation' accounts for the largest proportion of bugs associated with the `Incorrect Functionality' symptom, contributing 38.33\% of such cases.
This root cause leads to subtle logical defects in the underlying implementation, which can silently corrupt the model outputs.
The `Poor Performance' symptom is rare overall. It is related to two main categories (i.e., `API Misuse' and `Incorrect Algorithm Implementation'), which indicates that both incorrect API sequences and erroneous algorithms affect the throughput and efficiency of the LLM library.

\find{In the LLM library, crashes are mainly caused by API-level bugs, while functional errors mainly stem from algorithmic flaws.
This highlights the need for strict API validation and deep testing of the core logic of LLM libraries.}



\section{RQ3: Study of Existing Test Practices}\label{sec:rq3}


\begin{figure}[!tbp]
    \centering     
    \includegraphics[width=\linewidth]{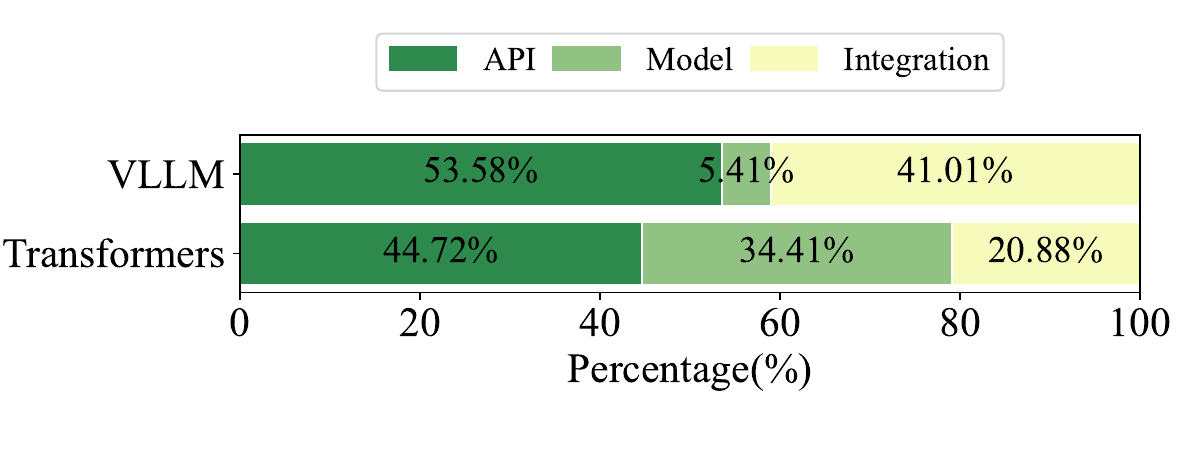}
    \caption{Test Distribution by Levels}
    \label{fig:test_level}
\end{figure}

\subsection{Test Level}\label{sec:rq3_testlevel}

LLM libraries can be tested at different levels of granularity, each serving distinct quality assurance purposes. 
We categorize tests into three levels in this work, as shown in~\autoref{fig:test_level}.

\textbf{API-Level Tests.}
LLM libraries consist of numerous foundational components that provide essential functionalities, such as tokenizers for text processing, attention mechanisms for sequence modeling, memory managers for KV cache allocation, and configuration parsers for model setup. API-level tests directly validate these individual components in isolation, rather than testing them indirectly through model execution or integration workflows. 
By testing components independently, API-level tests enable fast execution, precise fault localization, and comprehensive validation of the building blocks upon which all higher-level functionalities depend.
In DL testing, most work also focuses on API-level tests, e.g., Conv2D operators~\cite{deng2023largelanguagemodelszeroshot}.

\textbf{Model-Level Tests.}
These tests evaluate the functionality of complete model architectures as cohesive units, rather than isolated components. These tests typically involve executing the forward procedure on sample inputs and verifying that the outputs or hidden state values meet expected criteria.
While widely adopted in deep learning library testing, model-level tests face key limitations: they demand considerable computational resources, making them less scalable, and often provide limited coverage of lower-level components and APIs~\cite{zhang2024survey}.

\textbf{Integration Tests.}
Integration tests assess interactions across subsystems, validating end-to-end workflows such as raw text input through tokenization, inference, and output processing. 
For instance, such tests validate the complete execution of critical components like the \texttt{pipeline()} function or \texttt{Trainer} interface in Transformers.
Moreover, integration testing plays a pivotal role in verifying cross-compatibility across diverse model formats, distributed computing architectures, and external dependencies (e.g., PyTorch, TensorFlow). By simulating realistic, multi-component environments, these tests guarantee the robustness and reliability of the LLM library under complex, real-world usage scenarios.

We find that API-level tests constitute the largest proportion of test cases, accounting for 46.65\% overall (i.e., 53.58\% in vLLM and 44.72\% in Transformers), making it the most prevalent test category in both libraries.
Model-level tests make up 29.28\% of all cases. However, the distribution varies significantly between the two libraries: 34.41\% of test cases in Transformers are model-level tests, compared to only 5.41\% in vLLM. This discrepancy reflects the differing design goals of the two libraries—Transformers primarily focuses on model management and deployment, while vLLM is designed to leverage pre-trained models for efficient inference and performance optimization.
The remaining 24.07\% are integration tests, which validate interactions between components and ensure end-to-end functionality. Notably, vLLM devotes a much larger portion of its tests to this category (41.01\%) compared to Transformers (20.88\%), aligning with its emphasis on system-level performance and integration.

\find{
API-level testing is the most prevalent test strategy, accounting for 46.65\% of all cases. Due to differing design goals, Transformers emphasizes model-level testing (34.41\%), while vLLM focuses more on integration testing (41.01\%).
}

\subsection{Test Oracle}\label{sec:rq3_oracle}

Test oracles define how test outcomes are determined to be correct or incorrect. We identify several categories of oracles used in LLM library testing, each suited to different types of verification challenges.

\textbf{Pre-Defined Oracle}.
Developers often specify expected outputs for some given sample inputs, such as predefined floating-point tensors or text generation results. In scenarios involving type conversions or data format transformations, oracles may specify expected data types or shapes rather than exact values. 
We show an example that for a given input test, the expected output token id tensors are given:
\begin{lstlisting}[style=pythonstyle]
examples = [" Hello world", " DomDramg"]  # need leading spaces for equality
fairseq_results = [
    torch.tensor([0, 20920, 232, 2]),
    torch.tensor([0, 11349, 495, 4040, 571, 2])]
\end{lstlisting}

\textbf{Differential Testing}.
Differential testing validates functionality by comparing outputs from different implementations, versions, or configurations. We identify three categories: (1) \textit{Cross-device differential testing} compares outputs between different hardware platforms (e.g., CPU vs. GPU); (2) \textit{Intra-library differential testing} compares different internal APIs within the same library that should produce equivalent results; (3) \textit{Cross-library differential testing} compares outputs between different LLM libraries to ensure compatibility. 
This approach has been widely adopted in DL framework testing, with the most typical cross-framework backend consistency test~\cite{guo2020audee}.
For example, we find that vLLM often uses outputs from Transformers as reference oracles, comparing them against its own outputs to validate correctness.
\begin{lstlisting}[style=pythonstyle]
with hf_runner(model, dtype=dtype) as hf_model:
    hf_outputs = hf_model.generate_greedy_logprobs_limit(
        ...)

with vllm_runner(model, dtype=dtype) as vllm_model:
    vllm_outputs = vllm_model.generate_greedy_logprobs(
        ...)
check_logprobs_close( outputs_0_lst=hf_outputs, outputs_1_lst=vllm_outputs)
\end{lstlisting}

\textbf{Metamorphic Testing}.
Metamorphic testing generates test oracles by invoking the same code with systematically varied inputs and verifying that outputs satisfy expected relationships. We observe two subcategories: \textit{Invariant property testing} verifies that certain properties remain unchanged under specific transformations (e.g., tokenize-detokenize cycles should preserve original text), while \textit{relationship property testing} verifies that outputs change predictably when inputs are modified (e.g., increasing temperature parameters should increase randomness in text generation).

\textbf{Property-based Testing.}
Property-based testing is particularly useful in scenarios where specifying exact expected outputs is difficult or infeasible. Instead of asserting precise values, these tests verify that outputs adhere to certain expected properties or constraints. While this approach is less stringent than exact-match assertions, it is more scalable and often sufficient to catch a broad class of errors. For example, one may assert that the length of a generated sequence does not exceed a predefined \texttt{max\_length}, or that a model loading function does not return None—a simple yet effective validation that ensures the model was successfully instantiated.
\begin{lstlisting}[style=pythonstyle]
def test_model_from_pretrained(self):
    ...
    model = LukeModel.from_pretrained(model_name)
    self.assertIsNotNone(model)
\end{lstlisting}

\begin{figure}[!tbp]
    \centering     
    \includegraphics[width=\linewidth]{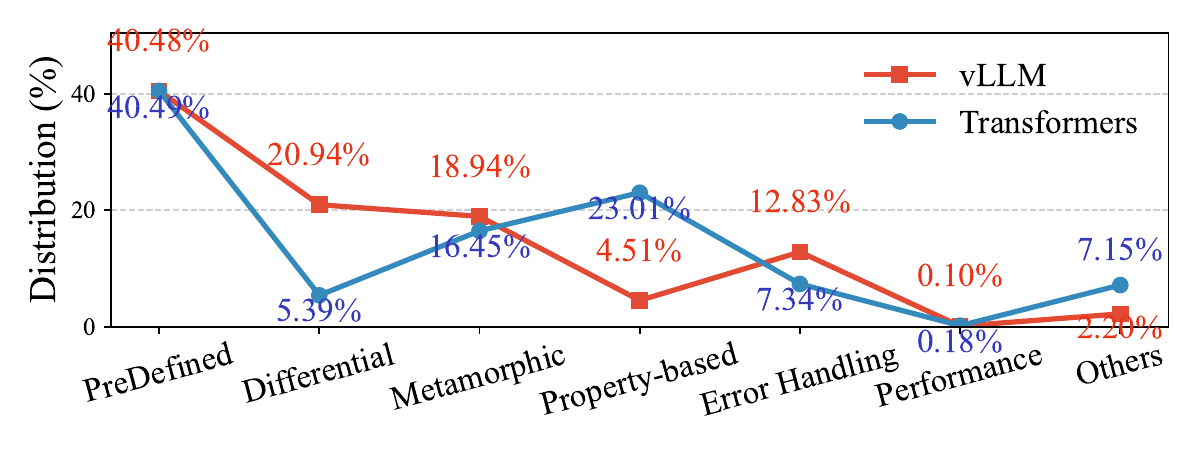}
    \caption{Test Distribution by Oracles}
    \label{fig:test_oracles}
    \vspace{-5pt}
\end{figure}

\textbf{Error Handling.}
Error handling tests verify that systems correctly detect and respond to invalid inputs and exceptional conditions. Many LLM libraries implement extensive precondition checking, raising appropriate exceptions for malformed, out-of-range, or incompatible inputs. These tests typically provide deliberately invalid inputs and verify that systems respond with appropriate error messages and exception types.
Robust error handling is crucial for LLM libraries because they often process user-provided data and configurations that may not conform to expected formats. Good error handling improves user experience by providing clear feedback when mistakes occur.

\textbf{Performance Testing.}
It focuses on verifying that systems meet specified resource utilization and efficiency requirements. 
Common scenarios include memory usage validation, where tests verify that operations stay within predefined bounds and respect parameters like \(low\_cpu\_mem\_usage\). Performance tests also examine computational efficiency, comparing execution times between different implementations and ensuring that optimizations produce expected improvements without introducing regressions.
Performance testing is particularly important for LLM libraries due to the computational intensity of large language models and the need to optimize resource usage for practical deployment.

\textbf{Others.}
This category includes other general-purpose checks that fall outside the scope of functionality or performance testing.
For instance, they may assert that initialization configurations are correctly specified, confirm the existence of required local directories or files.

As shown in~\autoref{fig:test_oracles}, pre-defined oracles are the most widely used in both libraries, accounting for 40.48\% of all test cases in Transformers and vLLM. This indicates a strong reliance on fixed expected outputs, particularly for unit-level functionalities and deterministic behaviors. However, the second most common oracle type differs between the two: Transformers primarily adopt property-based testing (23.01\%), while vLLM more frequently employs differential testing (20.94\%). This discrepancy reflects their design focuses—Transformers supports a wide range of models with inherently variable outputs, making property-based assertions more suitable; whereas vLLM, built atop Transformers, often uses its outputs as reference oracles, naturally lending itself to differential testing~\cite{vllmtest_2024}.
Nevertheless, this approach poses a risk that if the reference outputs in Transformers are erroneous, vLLM may inherit and propagate these errors, undermining the reliability of its test suite and overall correctness assurance.

\find{
Pre-defined oracles constitute the dominant testing approach (40.48\%) across both libraries. However, the secondary oracle strategies differ significantly: Transformers favors property-based testing, while vLLM relies more heavily on differential testing.
}

\begin{figure}[!tbp]
    \centering     
    \includegraphics[width=0.98\linewidth]{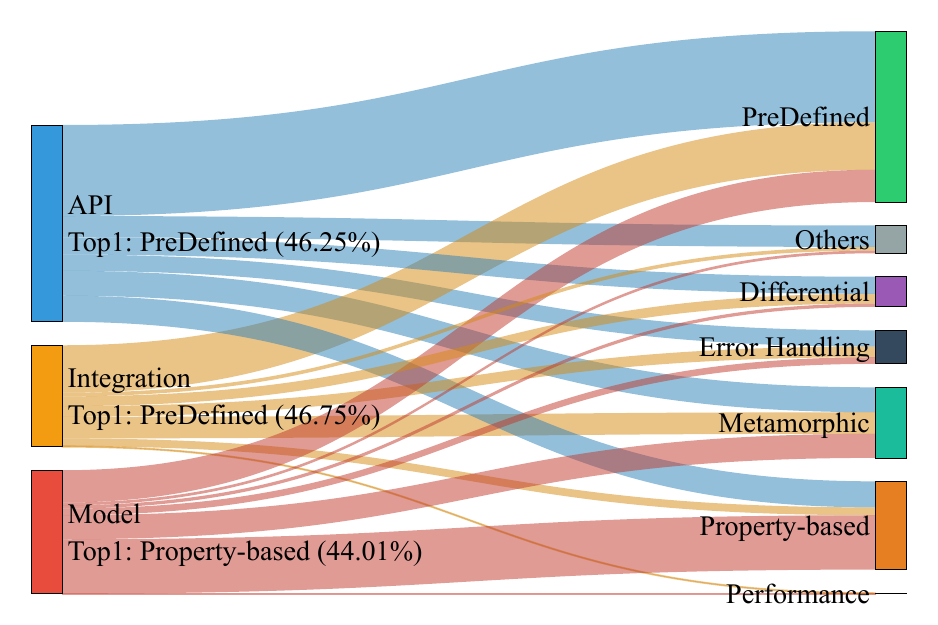}
    \caption{Distribution between Test Levels and Oracles}
    \label{fig:test_level_oracle}
    \vspace{-5pt}
\end{figure}

Moreover, as shown in~\autoref{fig:test_level_oracle}, we find that API-level and integration-level tests predominantly favor predefined oracles, whereas model-level tests exhibit a preference for property-based testing. This distinction reflects the inherent characteristics of each testing level: API-level and integration-level tests typically involve well-defined, end-to-end transformations (e.g., integration pipelines converting string inputs to string outputs), making predefined oracles more intuitive. In contrast, model-level testing requires validation of numerous hidden intermediate properties, rendering property-based testing more suitable.

\subsection{How about the effectiveness of existing tests}\label{sec:rq4}

Despite the existence of comprehensive test suites,  various bugs continue to be reported and fixed in LLM libraries, raising concerns about the actual effectiveness of current testing practices. To understand the limitations of existing test suites, we conduct an in-depth study focused on the Transformers library. We first analyze 230 bug-fixing commits and find that only 14 (6.10\%) are detected by test failures at the time of their introduction, while the majority remained undetected until later stages, indicating that existing tests often fail to identify regressions early.
To systematically investigate why these bugs evade detection, we collect 139 bugs that are compatible with our local testing environment spanning versions \texttt{v4.42.1} to \texttt{v4.47.1} for detailed execution analysis. We examine three critical questions: \ding{182} whether functions containing bugs are executed by the existing test suite, \ding{183} whether test cases trigger the buggy code snippets when these functions are executed, and \ding{184} whether test oracles successfully identify bugs when the buggy code is covered by test cases.


\textbf{Lack of Test Drivers.}
To evaluate whether the existing test suite is capable of invoking the buggy code, we extract all functions modified in the bug-related commits (referred to as buggy functions) and compare them with the set of functions executed by the current test suite (referred to as executed functions).
We find that 32.37\% of bugs involve functions that are never executed by any test, indicating a lack of appropriate test drivers. 
Since executing the buggy code is a necessary precondition for bug detection, these bugs remain undetectable regardless of oracle quality or test cases diversity.
This indicates that developers should pay more attention to writing tests for each newly added function, especially for those that are not directly invoked by other functions or APIs.

\textbf{Lack of Test Cases.}
Even when appropriate test drivers exist, many bugs escape detection due to inadequate test case diversity. Our analysis revealed that 41.73\% of bugs involve functions that are executed by existing tests but lack comprehensive scenario coverage, meaning tests failed to cover the specific code branches or execution paths that trigger the bugs.
This phenomenon is particularly prevalent in unit testing, where libraries often implement only basic input cases without exploring edge conditions or parameter variations. In model-level and integration testing scenarios, interfaces with dozens of parameters are frequently tested with only a single parameter combination. 
A representative example~\cite{transformers_pull33766} tests the \texttt{use\_cache} functionality in Idefics2's \(forward\) method, where existing tests only covered the \texttt{use\_cache=False} scenario, leaving the \texttt{use\_cache=True} branch completely untested and allowing related bugs to remain undetected.
The case highlights the need for more comprehensive test case design that considers diverse parameter combinations and execution paths to ensure effective bug detection.

\textbf{Lack of Test Oracles.}
More concerning, our analysis reveals that 25.90\% of bugs remain unidentified due to inadequate or incorrect test oracles, even when all lines of buggy code are executed by tests. This category represents bugs where the tests successfully reach the problematic code but fail to detect the resulting incorrect behavior.
It is important to note that test oracle limitations can co-exist with missing test drivers or cases. suggesting that the actual proportion of lack of oracle issues may be higher. 
As we discussed above, many tests in LLM libraries rely on indirect indicators such as tensor shapes or data types rather than validating semantic correctness. While these checks can catch structural errors, they often fail to detect subtle faults that alter model behavior without affecting superficial outputs.

\textbf{Example.} Consider the following buggy case in MBart's \texttt{resize\_token\_embedding} method, where embeddings are incorrectly cast from \texttt{MBartScaledWordEmbeddings} to standard \texttt{nn.Embeddings}, resulting in the loss of essential scaling attributes. The associated test case highlights two key oracle deficiencies: \ding{172} it verifies only superficial properties, such as embedding shape and basic I/O correctness, without checking whether the specific embedding type and its scaling behavior are preserved; and \ding{173} it uses simplistic inputs that fail to reveal the semantic consequences of losing the scaling mechanism on the model’s output. 

\begin{lstlisting}[style=pythonstyle]
def test_resize_tokens_embeddings_more(self):
    ...
    model.resize_token_embeddings(new_vocab_size)
    input_new, output_new = _get_embs(model)
    self.assertEqual(input_new.shape, (new_vocab_size, config.d_model))
    self.assertEqual(output_new.shape, (new_vocab_size, config.d_model))
    self.assertTrue(torch.eq(input_new, output_new).all())
    \\ Missing assert the type of the embeddings
\end{lstlisting}

\find{
Most bugs can not be identified by existing tests at the time of their introduction, revealing critical gaps in test effectiveness and adequacy. The most common causes are the lack of test cases (41.73\%), followed by missing test drivers (32.37\%) and insufficient oracles (25.90\%).
}


\section{DISCUSSIONS}\label{sec:discussions}

\subsection{Implications and Lessons Learned}\label{sec:implications}

\subsubsection{For LLM Library Developers}
We offer the following suggestions for developers:
\noindent\textbf{\ding{182} Mitigate API Misuse through rigorous design and validation.}
Our analysis reveals that API misuse is the dominant root cause, accounting for 32.17\%-48.19\% of bugs.
We recommend stricter interface specifications. Clear documentation and usage examples can further reduce API misuse incidents.
\noindent\textbf{\ding{183} Invest in more high-quality tests and oracles.}
Despite extensive test suites, our study reveals that most bugs escape detection due to missing test drivers (32.37\%), cases (41.73\%), and  oracles (25.90\%). 
LLM library developers should systematically extend test coverage, particularly for newly added components, and develop more effective oracles to detect semantic correctness issues beyond structural validation.

\subsubsection{For Users}
For LLM practitioners and downstream users, we highlight key practical implications:
\noindent\textbf{\ding{182} Validate Functional Correctness with Thorough Testing.}
Our findings show that incorrect functionality is the most prevalent and dangerous bug symptom, especially in vLLM (43.37\%). Unlike crashes, these silent failures often go unnoticed but can subtly corrupt model outputs over long sequences. Users should validate applications using multiple test inputs and longer sequences to expose subtle functional flaws.
\textbf{\ding{183} Actively Report Bugs to Improve Library Quality.}
Most bugs are discovered during real-world usage rather than caught by test suites, highlighting the essential role of community feedback. We encourage users to actively report issues with reproducible examples and environment details to help maintainers diagnose and fix problems, thereby enhancing both individual project reliability and the broader LLM ecosystem.

\subsubsection{For Researchers}
We strongly call on the research community to devote more attention to the reliability and testing of modern LLM libraries. 
Compared to traditional DL frameworks, LLM libraries present new and unique challenges in terms of complexity, modularity, and usage patterns. Based on our empirical findings, we outline several promising directions for future research:
\ding{182}\textbf{Automated Test Driver and Test Case Generation.} Our findings indicate that the lack of proper test drivers and test cases is the primary reason for undetected bugs. While existing LLM-based testing approaches show promise~\cite{deng2023largelanguagemodelszeroshot,deng2023large}, they face challenges with rapidly evolving APIs that outpace current LLM (which drives the testing tool) knowledge. Key research directions include leveraging LLMs to generate test drivers for emerging functionalities and developing lightweight, coverage-guided techniques to maximize rare branch testing while minimizing computational overhead.
\ding{183}\textbf{Advanced Test Oracle Development.} Current testing relies primarily on predefined outputs (poor scalability) and property-based oracles (high false negatives). Future work should explore automated oracle generation using LLMs, specification-based synthesis, and hybrid validation strategies to improve coverage and accuracy.


\subsection{Threat to Validity}\label{sec:threats}

\textbf{External Threat.} Our analysis is limited to two LLM libraries, which may not fully represent the entire LLM software ecosystem. To mitigate this, we selected HuggingFace Transformers (the most widely adopted foundational library) and vLLM (a representative high-performance inference library). Both libraries exhibit consistent patterns in testing practices and challenges, increasing confidence that results reflect broader LLM development trends. Additionally, our bug collection process may introduce bias; we followed established filtering keywords and inclusion criteria from prior work to ensure dataset reproducibility and relevance.

\textbf{Internal Validity.}
The internal threat to validity primarily concerns the bug and test case labeling process.
To mitigate this, we first adopted a taxonomy framework from existing literature to guide our initial annotation. 
Then, to minimize subjectivity, two authors independently labeled all samples and resolved disagreements through discussion, ensuring consistency and reliability in the classification process.

\section{RELATED WORK}\label{sec:rw}


Prior work has analyzed bug characteristics in general-purpose DL libraries like TensorFlow and PyTorch. Studies by Islam et al.\cite{10.1145/3338906.3338955} and Chen et al.\cite{chen2023toward} identified common symptoms (crashes, incorrect outputs) and root causes (numerical errors, API misuse). Specialized investigations examined silent faults~\cite{tambon2024silent} and performance bugs~\cite{cao2021characterizing}, inspiring tools like DeepPerf~\cite{cao2022understanding}. 
Research has also examined DL compilers~\cite{shen2021comprehensive,verma2022towards} and deployment environments including mobile~\cite{chen2020comprehensive} and JavaScript-based systems~\cite{ma2019moving,quan2022towards}.

Additionally, a limited body of work has examined testing practices in DL projects. For instance, Wang et al. investigated the usage of unit testing in open-source DL projects and its correlation with software quality~\cite{wang2024beyond}.
These empirical findings have motivated various DL testing techniques, including fuzzing~\cite{deng2023largelanguagemodelszeroshot,deng2023large,li2024seeds}, differential testing~\cite{10.1145/3510003.3510041}, coverage-guided testing~\cite{10.1109/ICSE48619.2023.00017}, and automated test generation~\cite{10.1145/3575693.3575707}. 
However, no prior work has systematically investigated the unique characteristics and testing challenges of modern LLM libraries.

In contrast, our work presents the first in-depth empirical study of modern LLM libraries, which differ significantly from traditional DL systems in architecture, abstraction level, and usage patterns. We reveal their unique bug characteristics and testing gaps, offering new insights to guide future testing research for LLM-centric infrastructures.

\section{CONCLUSION}\label{sec:conclusion}

In this work, we present the first comprehensive empirical study on bugs and testing practices in modern LLM libraries, analyzing 313 bug-fixing commits and over 7,700 test functions from two representative libraries. We systematically categorize bug symptoms, root causes, testing levels, and oracle strategies, revealing that most bugs evade detection due to inadequate test drivers, cases, and oracles. Based on our findings, we provide actionable recommendations for improving LLM library use, development, and testing.

\bibliographystyle{IEEEtran}
\bibliography{sample-base}

\end{document}